\documentclass[
aps,%
12pt,%
final,%
notitlepage,%
oneside,%
onecolumn,%
nobibnotes,%
nofootinbib,
superscriptaddress,%
noshowpacs,%
centertags]%
{revtex4}

\newcommand{\bp}{{\bf p}}

\def\qp{{q^\prime}}

\def\mn{m_N}

\def\bq{{\bf q}}
\def\bqp{{\bf \qp}}

\usepackage[utf8]{inputenc}
\usepackage[english,russian]{babel}
\usepackage{amsmath,amssymb}
\usepackage{graphicx}

\begin{document}

\title{On the contribution of the $P$ and $D$ partial-wave states  to the binding energy
of the triton    in the Bethe--Salpeter--Faddeev approach}

\author{\firstname{S.\,G.} \surname{Bondarenko}}
%
\author{\firstname{V.\,V.} \surname{Burov}}

\author{\firstname{S.\,A.} \surname{Yurev}}
\email[]{yurev@jinr.ru}
\affiliation{Bogoliubov  Laboratory of Theoretical Physics, JINR,
  Dubna, Russia}
\begin{abstract}
  The influence of the partial-wave states with nonzero orbital moment
of the nucleon pair on the binding energy of the triton in the relativistic case
is considered. The relativistic generalization of the Faddeev equation in the
Bethe--Salpeter formalism is applied.
Two-nucleon $t$ matrix is obtained from the Bethe--Salpeter equation with
separable kernel of nucleon--nucleon interaction of the rank one.
The kernel form factors are the relativistic type of the Yamaguchi functions.
The following two-nucleon partial-wave states are considered:
$^1S_0$, $^3S_1$, $^3D_1$, $^3P_0$, $^1P_1$, $^3P_1$.
The system of the integral equations are solved by using the iteration method.
The binding energy of the triton and three-nucleon amplitudes are found.
The contribution of the $P$ and $D$ states to the binding energy of triton is given.
\end{abstract}
\maketitle
\section{Introduction}

The study of three-nucleon systems has a long history and many works
is devoted to the description of such nuclei. One of the most common nonrelativistic descriptions 
is based on the application of the Faddeev equation with various two-particle
potentials.
Among such potentials, there are realistic ~[1] 
and separable~[2].
Such studies have made it possible to achieve
significant progress in the description of static
and dynamic properties of three-nucleon systems.

In the same time planned experiments on the scattering of electrons by $^3$He and
$^3$H, for instance, Jefferson Lab Experiment E1210103, with the energies of the initial
particles up to 12 GeV, require a relativistic description.
There are ways of relativizing the non-relativistic description,
and the methods that follow from the first principles of quantum field theory(QFT).
Among the latter we single out the quasipotential Gross equation with
the exchange kernel of a nucleon-nucleon interactions~[3],
and also a approaches based on the Bethe-Salpeter formalism with
 zero range of forces~[4], and 
with a separable kernel of  interaction ~[5,6].

This work develops the ideas represented in the articles~[5],
where the triton is considered in the $S$-state, and~[6],
where along with the $ S$-state, the contribution of the $ D$-state
into the two-particle $ t $ matrix was considered.
To describe a three-nucleon bound state, a relativistic
generalization of Faddeev equations in the Bethe-Salpeter formalism --
Bethe-Salpeter-Faddeev equation -- is used. For simplicity of calculations,
we consider nucleons have the same masses and the scalar propagators
instead of the spinor ones. The spin-isospin structure of the system is described
through matrices of recoupling coefficient  from one partial state to another.

In previous works~[7,8]
we considered the case of taking into account the $ D $-wave not only in the two-particle $ t $ matrix, 
but also its amplitudes in the system of integral equations.
In the present paper, the equation is generalized to the case of nonzero values of the angular momentum 
of a pair of nucleons ($L > 0: P$- и $D$-states).  The contributions of the following
two-particle partial states are considered 
(with a full   momentum  of two-nucleon system $j=0,1$): $^1S_0$, $^3S_1$, $^3D_1$, $^3P_0$, $^1P_1$, $^3P_1$.
The resulting system of 12 integral equations for real and imaginary parts of
amplitudes is solved by the iteration method and the binding energy of the triton,
as well as all three-particle amplitudes are finding.

The work is organized as follows: after a brief description of the solution
Bethe-Salpeter equations for two-nucleon states (sec. 2),
the relativistic  Bethe-Salpeter-Faddeev equation with scalar propagators is introdused,
and the partial-wave decomposition is performed (sec. 3).
In sec. 4 the results of solving the system of equations and their discussion is represent.

\section{Two particles case}

Since the kernel of the Faddeev equation, written in integral form,
contains a two-particle $ t $ matrix we first consider the two-body problem.

The system of two relativistic particles can be described using the Bethe-Salpeter equation.
Written for the two-particle $ t $ matrix, it has
the next view: 
\begin{eqnarray}
&&T(p,p';P) =  V(p,p';P) +\nonumber\\
&&+\frac{i}{(2\pi)^4}\int d^4k\, V(p,k;P)\, G(k;P)\, T(k,p';P),
\label{BS}
\end{eqnarray}
where $p = (p_1 - p_2)/2$ [$p' = (p'_1 - p'_2)/2$] --
the relative 4-momentum of the particles of the system
in the initial [final] state,
$s = P^2$ -- square of the total 4-momentum of the system $P = p_1 + p_2 =  p'_1 + p'_2$, 
$T(p,p';P)$ -- two-particle $ t $ matrix,
$V(p,k;P)$ -- kernel (potential) of a nucleon-nucleon ($ NN $) interaction,
$G(k;P)$ -- the product of two scalar propagators of nucleons,
\begin{eqnarray}
G^{-1}(k;P) = \big[(P/2 + k)^2 - \mn^2 + i\epsilon \big]
\big[(P/2 - k)^2 - \mn^2+ i\epsilon\big].
\label{prop}
\end{eqnarray}

Considering the equation~(\ref{BS}) in the center of mass of system of two particles
$P=(\sqrt{s},{\bf 0})$,
it is possible to separate the angular dependence and carry out partial-wave decomposition:
\begin{eqnarray}
&&T_{LL'}(p_0,|{\bf p}|,p_0',|{\bf p'}|;s) =  V_{LL'}(p_0,|{\bf p}|,p_0',|{\bf p'}|;s) +
\label{BS1}\\
&&+\frac{i}{4\pi^3}\int dk_0\, |{\bf k}|^2\, d|{\bf k}|
\sum_{L''}  V_{LL''}(p_0,|{\bf p}|,k_0,|{\bf k}|;s)\, G(k_0,|{\bf k}|;s)\, T_{L''L'}(k_0,|{\bf k}|,p_0',|{\bf p'}|;s),
\nonumber
\end{eqnarray}

In the present paper, for solving equation we use the kernel of the $ NN $-interaction
in the separable form (rank one):
\begin{equation}
V_{LL^{'}}(p_0,|{\bf p}|,p_0',|{\bf p'}|;s) =  \lambda g^{(L)}(p_0,|{\bf p}|) g^{(L')}(p_0',|{\bf p'}|).
\label{separ}
\end{equation}

If we substitute in equation~(\ref{BS1}) the kernel of the $ NN $-interaction
as~(\ref{separ}), then
the two-particle $ t $ matrix will also have a separable form:
\begin{equation}
T_{LL^{'}}(p_0,|{\bf p}|,p_0',|{\bf p'}|;s) =  \tau(s) g^{(L)}(p_0,|{\bf p}|) g^{(L')}(p_0',|{\bf p'}|),
\label{tsepar}
\end{equation}
where function $\tau$:
\begin{eqnarray}
\tau(s) = 1/(\lambda^{-1} + h(s))
\label{t06}
\end{eqnarray}
and
\begin{eqnarray}
&&h(s) = \sum_{L} h_{L}(s)= \nonumber \\
&&=-\frac{i}{4\pi^3}\, \int\, dk_0\,\int\, |{\bf k}|^2\, d|{\bf k}|\, 
\sum_{L} [g^{[L]}(k_0,|{\bf k}|)]^2 S(k_0,|{\bf k}|;s).
\label{h}
\end{eqnarray}

As form factors $g^{(L)}(p_0,|{\bf p}|)$ of kernel is used
relativistic generalization
 of the Yamaguchi-type functions~[9,10]
\begin{eqnarray}
g^{[S]}(p_0,|{\bf p}|) = \frac{1}{p_0^2-|{\bf p}|^2-\beta_0^2+i0},
\end{eqnarray}
\vskip -7mm
\begin{eqnarray}
  g^{[P]}(p_0,|{\bf p}|) = \frac{\sqrt{|-p_0^2+|{\bf p}|^2|}}
  {(p_0^2-|{\bf p}|^2-\beta_1^2+i0)^2},
\end{eqnarray}
\vskip -7mm
\begin{eqnarray}
  g^{[D]}(p_0,|{\bf p}|) = \frac{C_2(p_0^2-|{\bf p}|^2)}
  {(p_0^2-|{\bf p}|^2-\beta_2^2+i0)^2},
\end{eqnarray}
where $\lambda$, $\beta_0$, $\beta_1$, $\beta_2$ и $C_2$ --
the  parameters of the model, which are selected in this way
that the calculated values of the observed coincide with the corresponding
experimental data for them. As observable quantities
in this case can be taken the length and phase of the scattering, the effective radius,
and in the case when there is a bound state --
deuteron ($^3S_1 - ^3D_1$-state), -- binding energy.
Numerical values of parameters  $\lambda$ and $\beta$
can be found in~[11].


\section{Three particles case}

The system of three relativistic particles can be described using the Faddeev equations
in the Bethe-Salpeter formalism:
\begin{eqnarray}
 {\scriptsize
\Biggl[
\begin{array}{c}
T^{(1)}\\
T^{(2)}\\
T^{(3)}
\end{array}
\Biggr]
=
\Biggl[
\begin{array}{c}
T_{1}\\
T_{2}\\
T_{3}
\end{array}
\Biggr]
-
\Biggl[
\begin{array}{ccc}
0      & T_1G_1 & T_1G_1 \\
T_2G_2 & 0      & T_2G_2 \\
T_3G_3 & T_3G_3 & 0
\end{array}
\Biggr]
\Biggl[
\begin{array}{c}
T^{(1)}\\
T^{(2)}\\
T^{(3)}
\end{array}
\Biggr],
 }
 \label{eqn6}
\end{eqnarray}
where full $t$ matrix $T=\sum_{i=1}^3T^{(i)}$,  $G_i$ --
two-particle Green's function of 
particles $j$ и $n$ ($(ijn)$ obeys cyclic permutation): 
\begin{eqnarray}
G_i(k_j,k_n) = 1/(k_j^2-\mn^2+i\epsilon)/(k_n^2-\mn^2+i\epsilon),
\label{eqn7}
\end{eqnarray}
$T_i$ -- two-particle $ t $ matrix.

For a system of particles with the same masses, Jacobi variables can be introduced:
\begin{eqnarray}
  p_i = \frac12 (k_j-k_n),\,
  q_i = \frac13 K - k_i,\,
  K=k_1+k_2+k_3.
  \label{eqn9}  
\end{eqnarray}

On the basis of expression~(\ref{eqn9}) the equation~(\ref{eqn6}) can be rewritten
in the following way:
\begin{eqnarray}
  &&T^{(i)}(p_i,q_i;p_i',q_i';s) = (2\pi)^4 \delta^{(4)}(q_i-q_i') T_i(p_i;p_i';s)-
\label{eqn10}\\
&&-i\int\frac{d p_i''} {(2\pi)^4}
T_i(p_i;p_i'';s) G_i(k_j'',k_n'')   \times \nonumber\\
&&   \times\left[ T^{(j)} (p_j'',q_i'';p_i',q_i';s)
  + T^{(n)}(p_i'',q_i'';p_i',q_i';s)\right]
\nonumber
\end{eqnarray}

We introduce the amplitude $\Psi^{(i)}(p_i,q_i;s)$
for a bound three-particle state:
\begin{eqnarray}
\Psi^{(i)}(p_i,q_i;s) = \langle p_i,q_i | T^{(i)} | M_B \rangle \equiv 
\Psi_{L M}(p,q;s),
\label{eqn11}
\end{eqnarray}
where $M_B = \sqrt{s} = 3\mn-E_{B}$ -- mass of bound state (triton),
$s=K^2$ -- square of the total momentum.

To separation the angular integration and to implement partial-wave decomposition
it is need to account, that solution for the two-particle $ t $ matrix is found in the system of
the center of mass of two nucleons but the solution for the three-particle amplitude is sought in
the center of mass system of three nucleons. Since the radial functions
$g^{[L]}(q_0,|\bq|)$ depend on the square of the relative 4-momentum
the Lorentz transformation must be carried out only for arguments of
 spherical harmonics. In this paper we assume, that the components of the relative 4-vectors in the two systems coincide
  i.e  we omit the effects of the Lorentz transformation. In this case, the dependence of the three-particle
  amplitudes from two 4-vectors $p$ и $q$ can be divided .

We represent the total orbital angular momentum of a triton in the following form:
{\boldmath{${L} = {l} + {\lambda}$}},
where {\boldmath $l$} -- internal orbital angular momentum of a two-particle subsystem
and {\boldmath $\lambda$} -- orbital angular momentum of the third  particle relative to the two-particle subsystem.

In order to distinguish the explicit dependence of the amplitude on the angular momentum,
we will present it in the following form:
\begin{eqnarray}
  && \Psi_{L M}(p,q;s) = \sum_{a\lambda} \Psi^{(a)}_{\lambda L}(p_0,|\bp|,q_0,|\bq|;s)
  {\cal Y}^{(a)}_{\lambda LM}({\hat \bp},{\hat \bq}),
\label{eqn12a}\\
&& {\cal Y}^{(a)}_{\lambda LM}({\hat \bp},{\hat \bq}) =
\sum_{m\mu} C_{lm\lambda\mu}^{LM} Y_{lm}({\hat \bp}) Y_{\lambda\mu}({\hat \bq}),
  \nonumber
\end{eqnarray}
where two-nucleon states $a\equiv{^{2s+1}l_j}$ are characterized by $ s $ - spin,
$l$ -- angular and $ j $ total  momentum. In the equation~(\ref{eqn12a}) introduced
designation ${\hat {\bf a}} \equiv \Omega_{\bf a}$ for angular variables of 3-vector ${\bf a}$, $C$ -- the Clebsch-Gordan coefficients,
and $Y$ -- spherical functions.

Using the result of the previous section for the two-particle $ t $
matrix~(\ref{tsepar}) and after partial-wave decomposition  write the amplitude
$\Psi_{\lambda L}^{(a)}$ in a separable form:
\begin{eqnarray}
\Psi^{(a)}_{\lambda L}(p_0,|\bp|,q_0,|\bq|;s) = g^{(a)}(p_0,|\bp|)\times
\label{eqn12b}\\
\times\tau^{(a)}[(\frac{2}{3}\sqrt s+q_0)^2-\bq^{2}]\,
\Phi^{(a)}_{\lambda L}(q_0,|\bq|;s).
\nonumber
\end{eqnarray}
The functions $\Phi^{(a)}_{\lambda L}$ satisfy the following system
integral equations:
\begin{eqnarray}
  && \Phi^{(a)}_{\lambda L}(q_0,|\bq|;s) =
  \frac{i}{4\pi^3} \sum_{a'\lambda'}
  \int_{-\infty}^{\infty} dq_0'\int_{0}^{\infty} \bqp^{2}d|\bqp|\,
  Z^{(aa')}_{\lambda\lambda'}(q_0,q;q_0',|\bqp|;s)\times
\label{eqn15p}\\
&&\times\frac{\tau^{(a')}[(\frac{2}{3}\sqrt s+q_0')^2-\bqp^{2}]}
    {(\frac{1}{3}\sqrt s-q_0')^2-\bqp^{2}-m^2+i\epsilon}
\Phi^{(a')}_{\lambda'L}(q_0',|\bqp|;s),
\nonumber
\end{eqnarray}
with effective kernels
\begin{eqnarray}
 && Z^{(aa')}_{\lambda\lambda'}(q_0,|\bq|;q_0',|\bqp|;s) =
 C_{(aa')}\int d\cos\vartheta_{\bq\bqp}
 K^{(aa')}_{\lambda\lambda' L}(|\bq|,|\bqp|,\cos\vartheta_{\bq\bqp})\times
 \label{eqn16p}\\
 && \times\frac{
 g^{(a)}(-q_0/2-q'_0,|\bq/2+\bqp|)
 g^{(a')}(q_0+q'_0/2,|\bq+\bqp/2|)}
 {(\frac{1}{3}\sqrt s +q_0+q'_0)^2-(\bq+\bqp)^2-\mn^2+i\epsilon},
 \nonumber
\end{eqnarray}
where
\begin{eqnarray}
&&K^{(aa')}_{\lambda\lambda' L}(|\bq|,|\bqp|,\cos\vartheta_{\bq\bqp}) =
(4\pi)^{3/2}\frac{\sqrt{2\lambda + 1}}{2L+1}\times
  \label{eqn17p}\\
&&
\times\sum_{mm'}C^{L m}_{l m \lambda 0}
C^{L m}_{l' m' \lambda' m-m'}
{Y}_{l m}^{*}(\vartheta,0) 
{Y}_{l' m'}(\vartheta',0)
{Y}_{\lambda' m-m'}(\vartheta_{\bq\bqp},0)
\nonumber
\end{eqnarray}
and
\begin{eqnarray}
  \cos\vartheta=(\frac{|\bq|}{2} + |\bqp|\cos\vartheta_{\bq\bqp})/
  |\frac{\bq}{2}+\bqp|, \nonumber \\
  \quad\quad\cos\vartheta'=(|\bq|+\frac{|\bqp|}{2}\cos\vartheta_{\bq\bqp})/
  |\bq+\frac{\bqp}{2}|.
\nonumber
\end{eqnarray}
The details of the calculation of the function $ K $ can be found in~[12].

Since we are considering the ground state of a three-nucleon system  $L=0$
and correspondingly
$l=\lambda, l'=\lambda'$, and the function $ K $ can be rewritten in the following form:
\begin{eqnarray}
&&K^{(aa')}_{ll 0} = \sqrt{(4\pi)^3} {Y}_{l0}^*(\vartheta,0) A_l'(\vartheta',\vartheta_{\bq\bqp}), \nonumber\\  
&&A_{l}(\vartheta',\vartheta_{\bq\bqp}) = \sum_{m'}
  C^{0 0}_{l m' l -m'} {Y}_{l m'}(\vartheta',0)
  {Y}_{l  -m'}(\vartheta_{\bq\bqp},0),\nonumber
\end{eqnarray}
where $l,l'$ correspond to the orbital moments of the partial states $[a,a']$. 

The accounting of spin-isospin structure of the equation kernel can be expressed in terms of matrix of recoupling coefficients from one partial state to another
[$(a)=^1S_0,^3S_1,^3D_1,^3P_0,^1P_1,^3P_1$],
which have the following form:
\begin{eqnarray}
C_{(aa')}
= \frac{1}{4}
\begin{pmatrix}
1   & -3  & -3 & \sqrt 3 & -\sqrt 3 & \sqrt 3 \\
-3  & 1  & 1 &\sqrt 3 & -\sqrt 3 & \sqrt 3 \\
-3  & 1  & 1 & \sqrt 3 & -\sqrt 3 & \sqrt 3 \\
\sqrt 3  & \sqrt 3  & \sqrt 3 & -1 &  -3 & -1 \\
-\sqrt 3  & -\sqrt 3  & -\sqrt 3 & -3 & -1 & -3 \\
\sqrt 3  & \sqrt 3  & \sqrt 3 & -1 &  -3 & -1 \\
\end{pmatrix}.
\label{eqn15a}
\end{eqnarray}

The system of integral equations~(\ref{eqn15p})--(\ref{eqn17p}) has singularities
however, in the case of a system of three coupled particles ($\sqrt s < 3\mn$)
all these singularities do not intersect the path of integration over $q_0$ and
thus do not affect the implementation of the procedure of Wick rotation
$q_0 \rightarrow  iq_4$.

System~(\ref{eqn15p})--(\ref{eqn17p}) after Vick rotation can be solved
using standard methods for solving integral equations.
One of them is discussed in the next section.

\section{Numerical calculations and results}

In this paper, a homogeneous system of 12 integral
equation with a parameter, which is the binding energy  of the  triton,
was solved using the iteration method.
A homogeneous system of integral equations has a solution not for all
values of parameter,
but only for those that satisfy some properties.

To determine the binding energy, the following condition was used
(in more detail~[13]):
\begin{eqnarray}
\lim_{n \to \infty}\frac{\Phi_n(s)}{\Phi_{n-1}(s)}\Big|_{s=M_B^2} = 1,
\end{eqnarray}
where  $n$ -- iteration number.

The procedure for solving the system of integral equations ~(\ref{eqn15p})--(\ref{eqn17p}) 
by the iteration method has good convergence. In numerical calculations of the binding energy of triton 
 and the amplitudes of its states for the Yamaguchi potential, 
the ratio of the previous  iteration to the next did not change with the growth of the iteration number
up to the sixth decimal place starting with the 20th iteration. 

For the numerical calculation of the integrals, the Gauss method on a two-dimensional
grid of nodes by dimension $N_1 \times N_2$ was used  with mapping
$q_4 = (1+x)/(1-x), |\bq| = (1+y)/(1-y)$.
The influence of the number of nodes on the convergence of the result of numerical integration  was investigated.
For integration on $|\bq|$ was enough $N_2 = 15$ nodes. With further increase quantity of nodes 
the numerical value of the integral did not change any more.
For integration on $q_4$ it was not enough such quantity of nodes.
 For the study of convergence we increased  quantity of nodes to $N_1=96$.
  With further increase quantity of nodes 
 the numerical value of the integral  changed only in the fourth decimal place.
This accuracy is sufficient and
allows us to take into account the contribution of various states to the binding energy.

The table represented  the calculated values of the binding energy
 for different probabilities of the $D$-state ($p_D = 4,5,6$).

\begin{table}[ht]
{The values of the binding energy of a triton (MeV)}
\label{tab1}
\begin{center}
\begin{tabular}{c|c|c|c|c|c}
\hline
$p_D$ & $^1S_0 - ^3S_1$ & $^3D_1$ & $^3P_0$ & $^1P_1$ & $^3P_1$ \\
\hline  
4     & 9.221          & 9.294  &  9.314  &  9.287 &  9.271 \\
5     & 8.819	       & 8.909 	&  8.928  &  8.903 &  8.889 \\
6     & 8.442	       & 8.545	&  8.562  &  8.540 &  8.527 \\
\hline
\multicolumn{6}{c}{Experiment \qquad\qquad 8.48} \\
\hline
\end{tabular}
\end{center}
\end{table}

The above results show that the main contribution to the binding energy of the triton
give an $ S $-state. Contribution of  $D$-state is positive and varies
from $0.8$ to $1.2$ \%
depending on the probability of $D$-state in deuteron ($p_D$ = 4-6 $\%$).
Contributions of $P$-states have different signs and partially compensate each other,
and their total contribution is $-0.2 \%$. So total contribution
two-particle  $P$- and $D$-partial states  with a full angular momenta $j=0,1$
into the binding energy of a triton is from $0.5$ to $1$ \%.
Comparison of nonrelativistic and relativistic calculations of binding energy 
was conducted in~[5].
The paper shows that relativistic calculation of binding energy in the case of accounting only 
 $S$-states more nonrelativistic at  $0.44$ MeV.

\begin{figure}[ht]
\center{
	\begin{tabular}{cc}
	\includegraphics[width=0.35\linewidth,angle=-90]{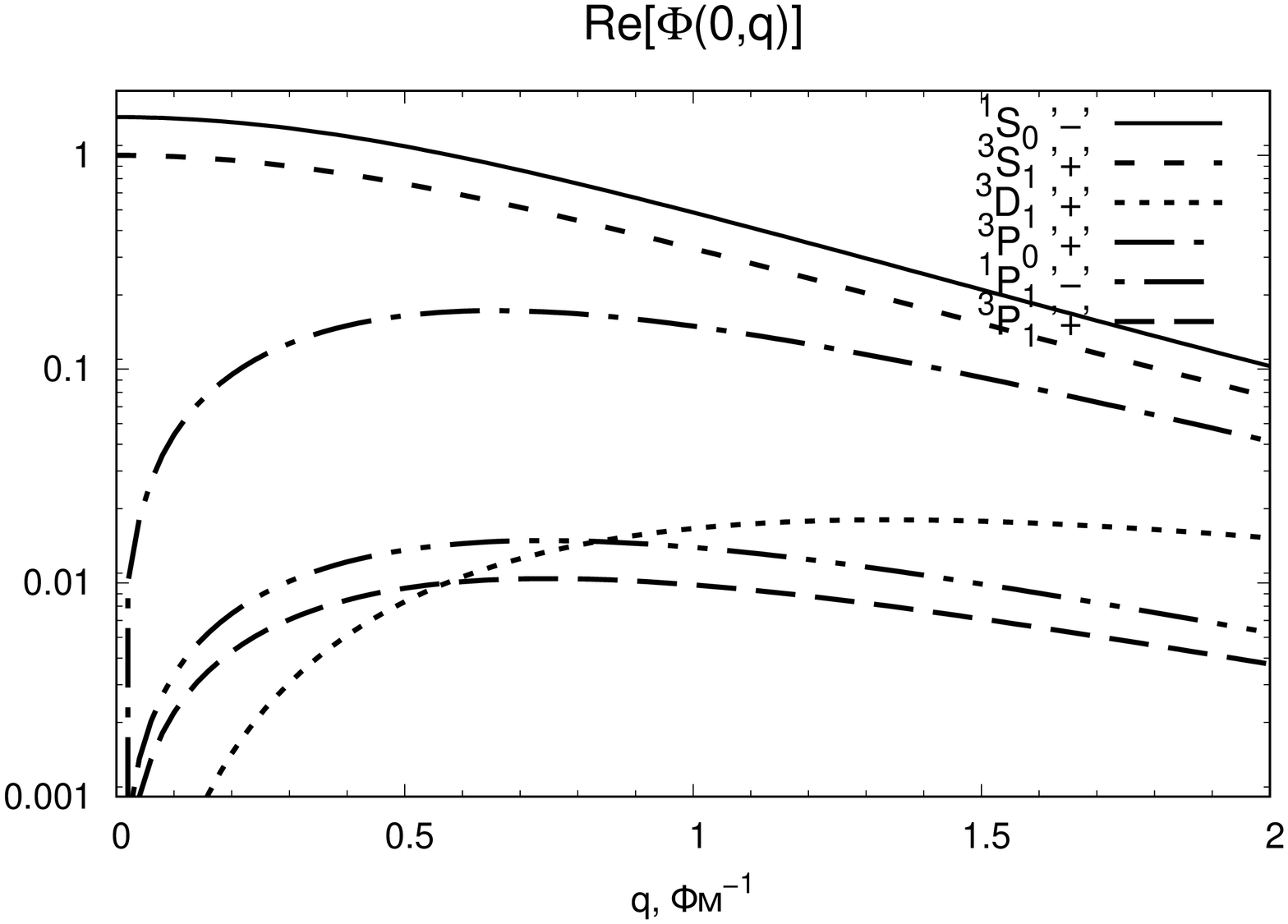}&
	\includegraphics[width=0.35\linewidth,angle=-90]{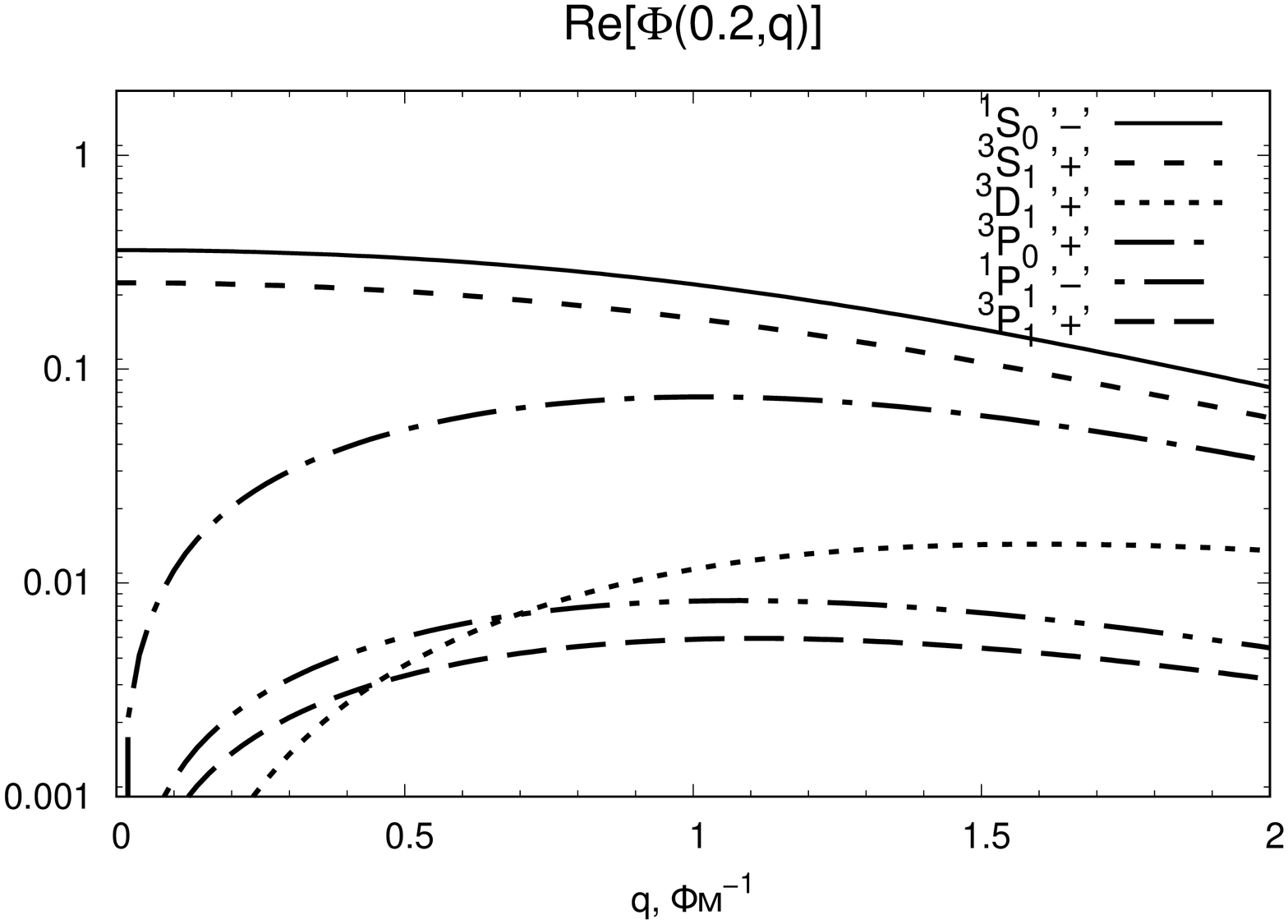}
	\end{tabular}
}
\caption{The real part of the amplitudes for all states considered in the work as a function of |\bq| with the value $q_4$ = 0 and $q_4$ = 0.2 Fm$^{-1}$.}
\label{1}
\end{figure}

\begin{figure}[ht]
\center{
	\begin{tabular}{cc}
	\includegraphics[width=0.35\linewidth,angle=-90]{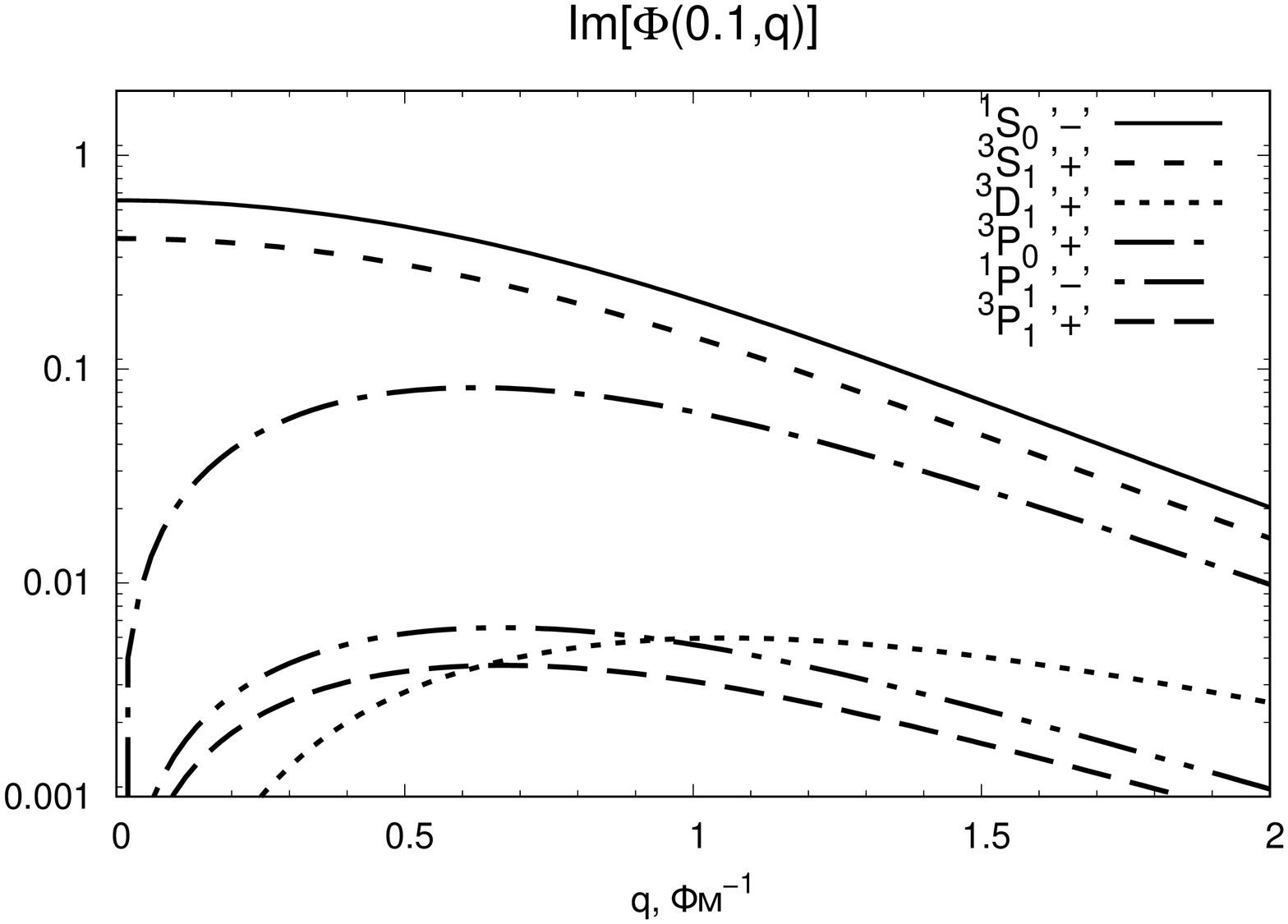}&
	\includegraphics[width=0.35\linewidth,angle=-90]{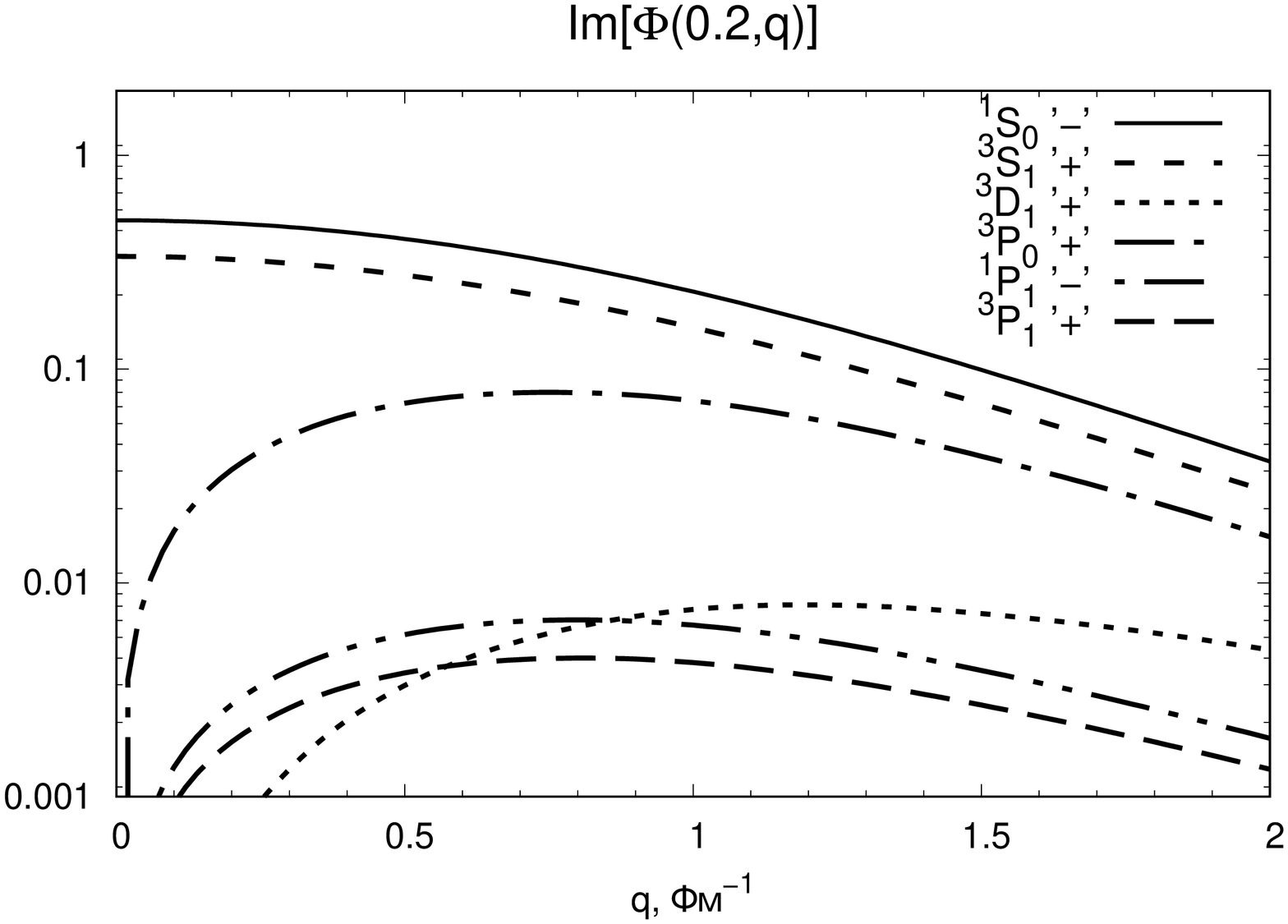}
	\end{tabular}
}
\caption{ The imaginary part of the amplitudes for all states considered in the work as a function of |\bq| with the value $q_4$ = 0.1 and $q_4$ = 0.2 Fm$^{-1}$.}
\label{2}
\end{figure}

\begin{figure}[ht]
\center{
	\begin{tabular}{cc}
	\includegraphics[width=0.35\linewidth,angle=-90]{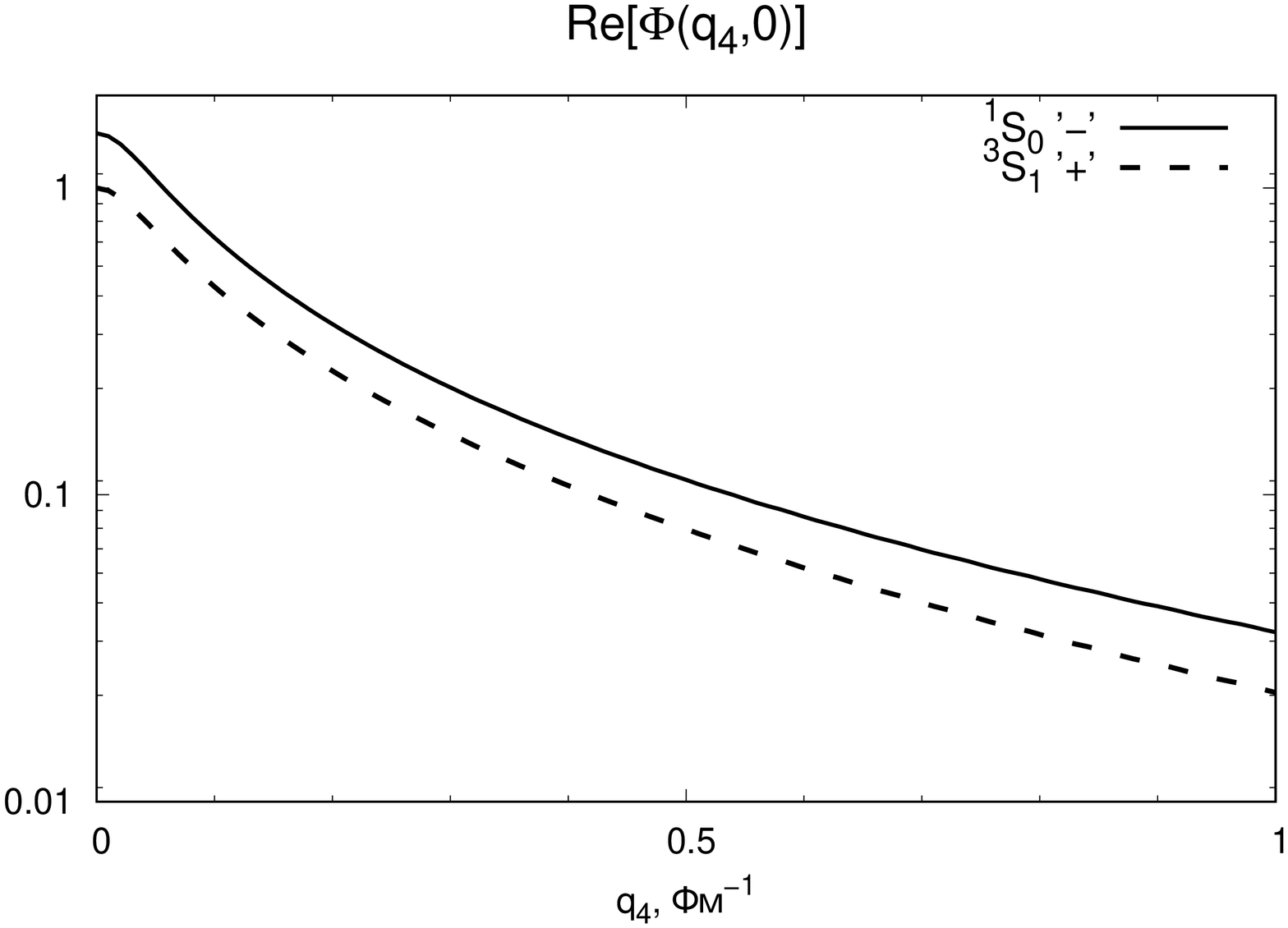}&
	\includegraphics[width=0.35\linewidth,angle=-90]{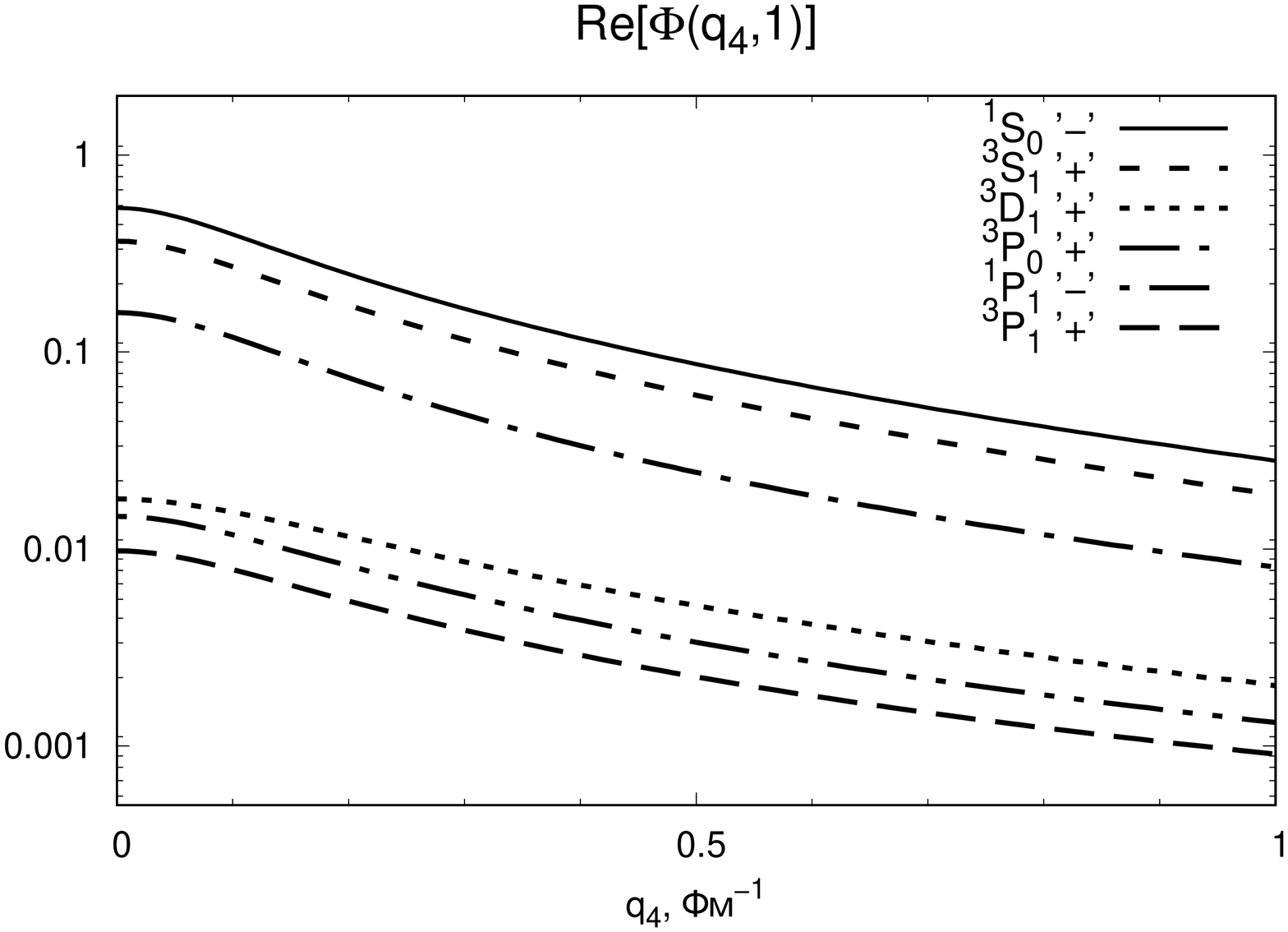}
	\end{tabular}
}
\caption{The real part of the amplitudes for all states considered in the work as a function of $q_4$ with the value |\bq| = 0 and |\bq| = 1Fm$^{-1}$.}
\label{3}
\end{figure}

\begin{figure}[ht]
\center{
	\begin{tabular}{cc}
	\includegraphics[width=0.35\linewidth,angle=-90]{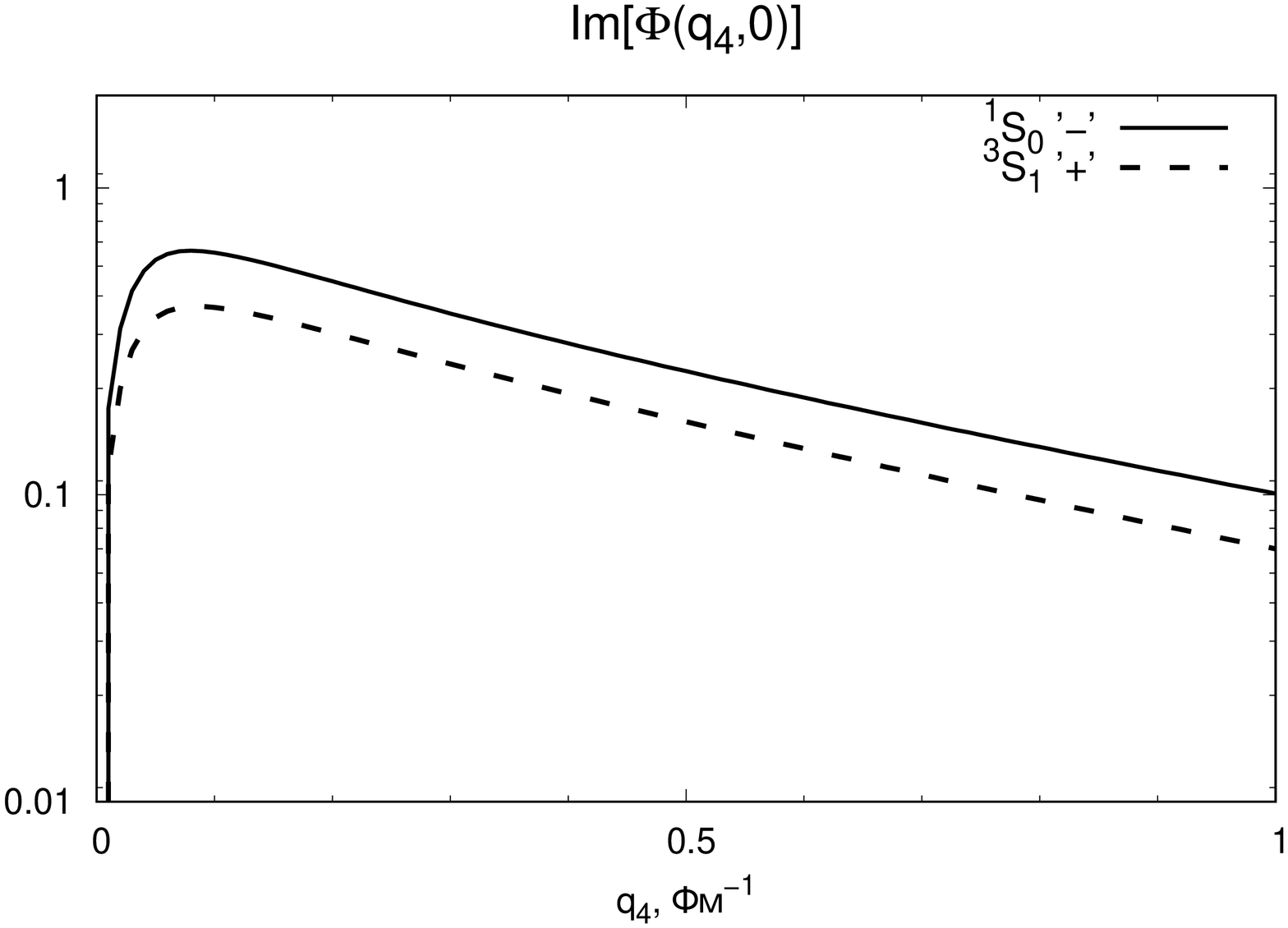}&
	\includegraphics[width=0.35\linewidth,angle=-90]{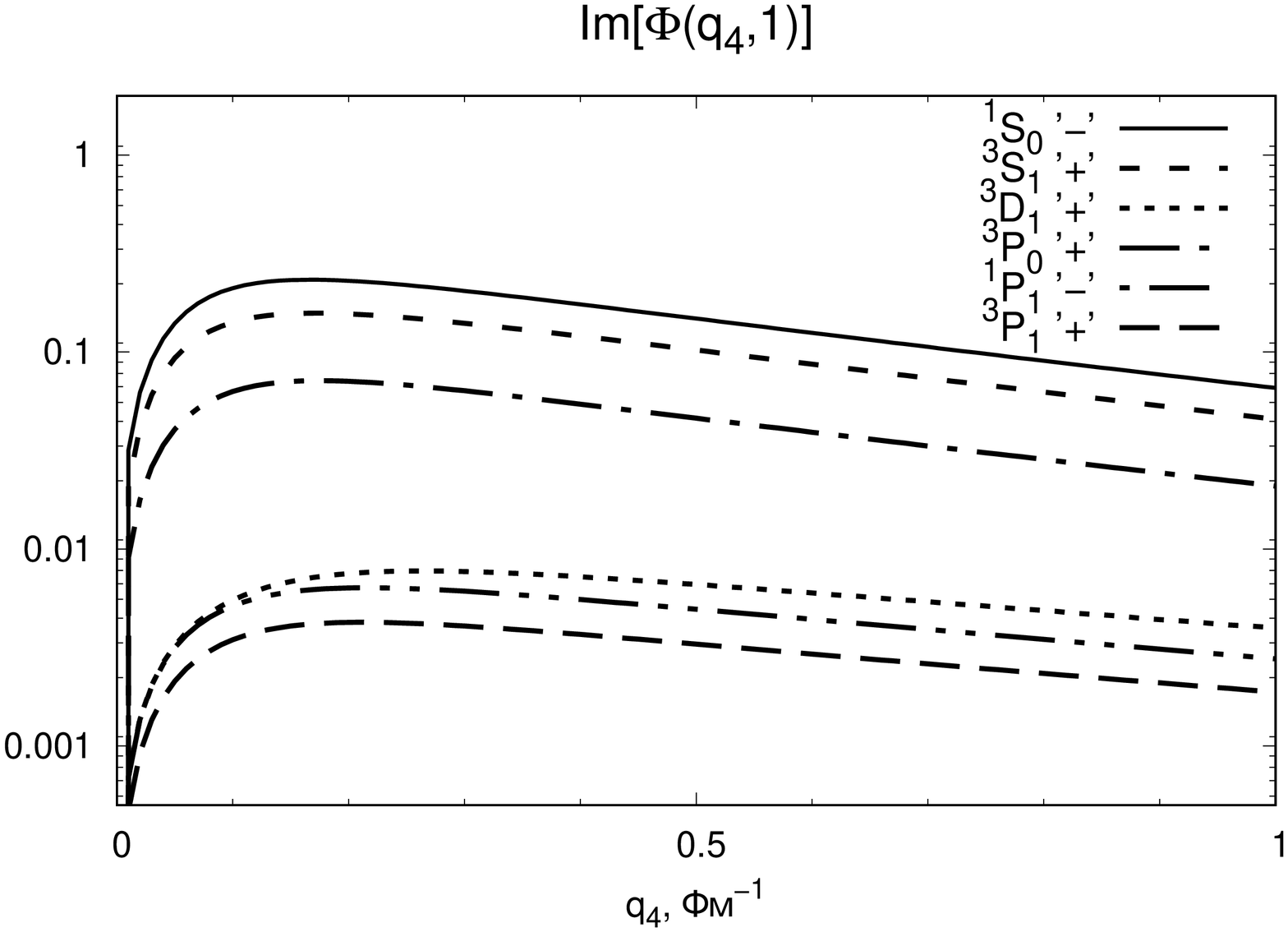}
	\end{tabular}
}
\caption{The imaginary part of the amplitudes for all states considered in the work as a function of $q_4$  with the value  |\bq| = 0 and |\bq| = 1 Fm$^{-1}$.}
\label{4}
\end{figure}

In Fig.~\ref{1}-\ref{4} represented graphs of real and imaginary parts of partial amplitudes on variables $|\bq|$
(at fixed values of $q_4$) and $q_4$
(at fixed values of $|\bq|$).
As can be seen from the graphs, amplitudes of  $S$-states dominate
wherein other states
give a nonzero contribution. However we believe that interference contributions of
 $S$-, $P$- and $D$-states in form factors of the three-particle system
must be taken into account in calculations.
Obtained amplitudes will be used to calculate the electromagnetic form factors of the triton
using the approximations described in articles ~[5,6].


\section{Conclusion}
The solution of the relativistic 
Bethe - Salpeter - Faddeev equation for a three-nucleon system (triton) are considered in article.
A relativistic generalization of the partial-wave decomposition procedure is carried out,
which is spread to the nonzero orbital angular momenta of the interacting pair of nucleons.
The case of $ S $ -, $ P $ - and $D$-partial states of the two-particle
subsystems are considered. The using of partial-wave decomposition and potential of
$NN$-interactions in a separable form led to a system of integral
equations for the amplitudes of states with different orbital
moments of particles in the nucleus.
The numerical solution of this system using the iteration method allowed
find the binding energy of a triton
and  amplitudes of its $ ^ 1S_0 $, $ ^ 3S_1 $, $ ^ 3D_1 $, $ ^ 3P_0 $, $ ^ 1P_1 $, $ ^ 3P_1 $ -states
as functions of two variables. 

This work was partially supported by the Russian Foundation for Basic Research grants \textnumero 16-02-00898  and \textnumero 18-32-00278.

\newpage	
\section*{References}
\begin{enumerate}

\item
  E.~van Faassen and J.~A.~Tjon,
  Phys.\ Rev.\ C {\bf 33}, 2105 (1986).

\item
  G.~Rupp, L.~Streit and J.~A.~Tjon,
  Phys.\ Rev.\ C {\bf 31}, 2285 (1985).

\item
  A.~Stadler, F.~Gross and M.~Frank,
  Phys.\ Rev.\ C {\bf 56}, 2396 (1997).

\item
  E.~Ydrefors, J.~H.~Alvarenga Nogueira, V.~Gigante, T.~Frederico and V.~A.~Karmanov,
  Phys.\ Lett.\ B {\bf 770}, 131 (2017).
  
\item
  G.~Rupp and J.~A.~Tjon,
  Phys.\ Rev.\ C {\bf 37}, 1729 (1988).

\item
  G.~Rupp and J.~A.~Tjon,
  Phys.\ Rev.\ C {\bf 45}, 2133 (1992).

\item
  S.~G.~Bondarenko, V.~V.~Burov and S.~A.~Yurev,
  EPJ Web Conf.\  {\bf 108}, 02015 (2016).

\item
  S.~Bondarenko, V.~Burov and S.~Yurev,
  EPJ Web Conf.\  {\bf 138}, 06003 (2017).

\item
  Y.~Yamaguchi,
  Phys.\ Rev.\  {\bf 95}, 1628 (1954).

\item
  Y.~Yamaguchi and Y.~Yamaguchi,
  Phys.\ Rev.\  {\bf 95}, 1635 (1954).
  
\item
  S.~G.~Bondarenko, V.~V.~Burov and S.~A.~Yurev,
  Phys.Part.Nucl.Lett. {\bf 15}, 442 (2018).

\item
  A.~Ahmadzadeh and J.~A.~Tjon,
  Phys.\ Rev.\  {\bf 139}, B1085 (1965).

\item
  R.~A.~Malfliet and J.~A.~Tjon,
  Nucl.\ Phys.\ A {\bf 127}, 161 (1969).
  
	
\end{enumerate}


\clearpage

\end{document}